\documentclass[pdflatex,iicol]{sn-jnl}

\usepackage[super,comma,sort]{natbib}

\jyear{2021}%

\graphicspath{{figs/}}

\begin{document}

\title[~]{X-ray detection of a nova in the fireball phase}

\author*[1]{\fnm{Ole} \sur{K\"onig}}\email{ole.koenig@fau.de}
\author*[1]{\fnm{J\"orn} \sur{Wilms}}\email{joern.wilms@sternwarte.uni-erlangen.de}
\author[2]{\fnm{Riccardo} \sur{Arcodia}}%\email{arcodia@mpe.mpg.de}
\author[1]{\fnm{Thomas} \sur{Dauser}}%\email{thomas.dauser@sternwarte.uni-erlangen.de}
\author[2]{\fnm{Konrad} \sur{Dennerl}}%\email{kod@mpe.mpg.de}
\author[3]{\fnm{Victor} \sur{Doroshenko}}%\email{doroshv@astro.uni-tuebingen.de}
\author[2]{\fnm{Frank} \sur{Haberl}}%\email{fwh@mpe.mpg.de}
\author[1]{\fnm{Steven} \sur{H\"ammerich}}%\email{steven.haemmerich@fau.de}
\author[1]{\fnm{Christian} \sur{Kirsch}}%\email{christian.ck.kirsch@fau.de}
\author[1]{\fnm{Ingo} \sur{Kreykenbohm}}%\email{Ingo.Kreykenbohm@fau.de}
\author[1]{\fnm{Maximilian} \sur{Lorenz}}%\email{maximilian.ml.lorenz@fau.de}
\author[2]{\fnm{Adam} \sur{Malyali}}%\email{amalyali@mpe.mpg.de}
\author[2]{\fnm{Andrea} \sur{Merloni}}%\email{am@mpe.mpg.de}
\author[2]{\fnm{Arne} \sur{Rau}}%\email{arau@mpe.mpg.de}
\author[3]{\fnm{Thomas} \sur{Rauch}}%\email{thomas.rauch@uni-tuebingen.de}
\author[4,5]{\fnm{Gloria} \sur{Sala}}%\email{gloria.sala@upc.edu}
\author[6]{\fnm{Axel} \sur{Schwope}}%\email{aschwope@aip.de}
\author[3]{\fnm{Valery} \sur{Suleimanov}}%\email{suleimanov@astro.uni-tuebingen.de}
\author[1]{\fnm{Philipp} \sur{Weber}}%\email{philipp.ph.weber@fau.de}
\author[3]{\fnm{Klaus} \sur{Werner}}%\email{werner@astro.uni-tuebingen.de}

\affil*[1]{\orgdiv{Dr.~Karl Remeis-Observatory and Erlangen Centre for
    Astroparticle Physics}, \orgname{Friedrich-Alexander-Universit\"at
    Erlangen-N\"urnberg}, \orgaddress{\street{Sternwartstr.~7},
    \postcode{96049} \city{Bamberg}, \country{Germany}}}

\affil[2]{\orgname{Max-Planck-Institut f\"ur extraterrestrische
    Physik}, \orgaddress{\street{Gie{\ss}enbachstra\ss{}e~1},
    \postcode{85748} \city{Garching}, \country{Germany}}}

\affil[3]{\orgdiv{Institut f\"ur Astronomie und Astrophysik, Kepler
    Center for Astro and Particle Physics},
  \orgname{Eberhard Karls Universit\"at}, \orgaddress{\street{Sand~1},
    \postcode{72076} \city{T\"ubingen}, \country{Germany}}}

\affil[4]{\orgdiv{Departament de F\'isica EEBE}, \orgname{Universitat Polit\'ecnica de Catalunya (UPC)}, \orgaddress{\street{Av. Eduard Maristany~10–14}, \city{Barcelona}, \postcode{08019}, \country{Spain}}}

\affil[5]{\orgname{Institut d'Estudis Espacials de Catalunya (IEEC)}, \orgaddress{\street{Carrer del Gran Capit\`a~2}, \city{Barcelona}, \postcode{08034}, \country{Spain}}}

\affil[6]{\orgname{Leibniz-Institut f\"ur Astrophysik Potsdam},
  \orgaddress{\street{An der Sternwarte~16},
    \postcode{14482} \city{Potsdam},  \country{Germany}}}


\abstract{
% ~200 words
\textbf{Novae are caused by runaway thermonuclear burning in the
  hydrogen-rich envelopes of accreting white dwarfs, which results in
  the envelope to expand rapidly and to eject most of its
  mass\citep{Chomiuk21a,starrfield16a}. For more than 30 years, nova
  theory has predicted the existence of a ``fireball'' phase following
  directly the runaway fusion, which should be observable as a short,
  bright, and soft X-ray flash before the nova becomes visible in the
  optical\citep{hillman14,Starrfield90a,Krautter08a}. Here we present
  the unequivocal detection of an extremely bright and very soft X-ray
  flash of the classical Galactic nova YZ~Reticuli 11\,hours prior to
  its 9\,mag optical brightening. No X-ray source was detected
  4\,hours before and after the event, constraining the duration of
  the flash to shorter than 8\,hours. In agreement with theoretical
  predictions\citep{Starrfield90a,Kato15a,Kato16a,Morii16a}, the
  source's spectral shape is consistent with a black body of
  $3.27^{+0.11}_{-0.33}\times 10^5$\,K ($28.2^{+0.9}_{-2.8}$\,eV), or
  a white dwarf atmosphere, radiating at the Eddington luminosity,
  with a photosphere that is only slightly larger than a typical white
  dwarf. This detection of the expanding white dwarf photosphere
  before the ejection of the envelope provides the last link of the
  predicted photospheric lightcurve evolution and opens a new window
  to measure the total nova energetics.}
}

\keywords{novae, cataclysmic variables, stars: individual (YZ~Reticuli), white dwarfs, X-rays: binaries}

\maketitle

\section*{eROSITA detection of YZ~Reticuli}

\begin{figure}[h]
  \centering
  \includegraphics[width=\linewidth]{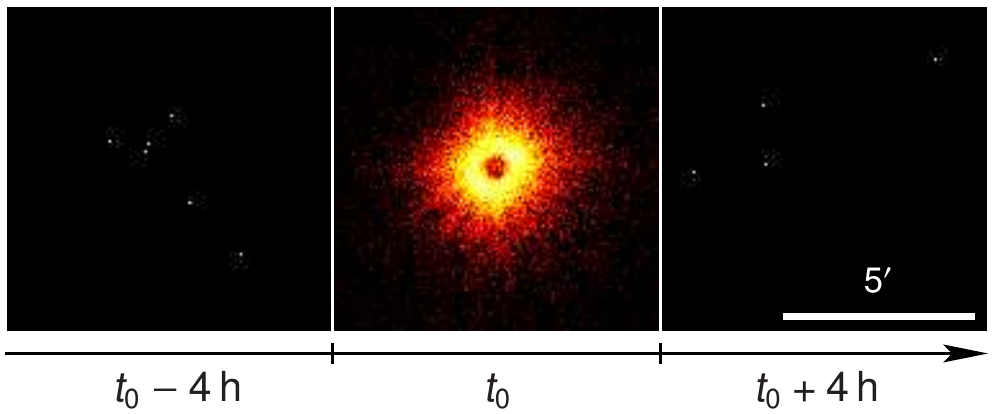}
  \caption{Sky images of all seven eROSITA cameras combined
    (0.2--0.6\,keV). On $t_0=2020$-07-07 16:47:20.64 TT, during the
    second all-sky survey, eROSITA detected a bright, new, soft X-ray
    flash that was severely affected by pile-up. No source can be seen
    in the scans four hours before and after the event.}
  \label{fig:flash_images}
\end{figure}

During its second all-sky survey (2020-06-11 -- 2020-12-15), the
eROSITA instrument\citep{Predehl21a} on board Spectrum-Roentgen-Gamma
(SRG)\citep{Sunyaev21a} scanned the field around
$\alpha_\mathrm{J2000.0}= 03\mathrm{h}\,58\mathrm{m}\,30\mathrm{s}$,
$\delta_\mathrm{J2000.0} = -54^\circ \,46'\,41''$ twenty-eight times,
separated by 4\,hours each. No source was detected for the first 22
scans. During the 23rd passage over the position, which started on
$t_0=2020$-07-07 16:47:20 TT, a new and extremely bright source was
detected (Fig.~\ref{fig:flash_images}). No source was visible in the
subsequent scans, constraining the event's duration to
$<8\,\mathrm{h}$. The position coincides with the location of Nova
YZ~Reticuli (= EC~03572$-$5455 = Nova Reticuli 2020), for which an
optical outburst was discovered\citep{McNaughtCBET4811} on 2020-07-15
14:09 UT. Subsequently, the object was classified as a classical He/N
Galactic nova\citep{Aydi20a,Carr20a} with a geometric distance of
$2.53^{+0.25}_{-0.26}$\,kpc.

\begin{figure}
  \centering
  \includegraphics[width=\linewidth]{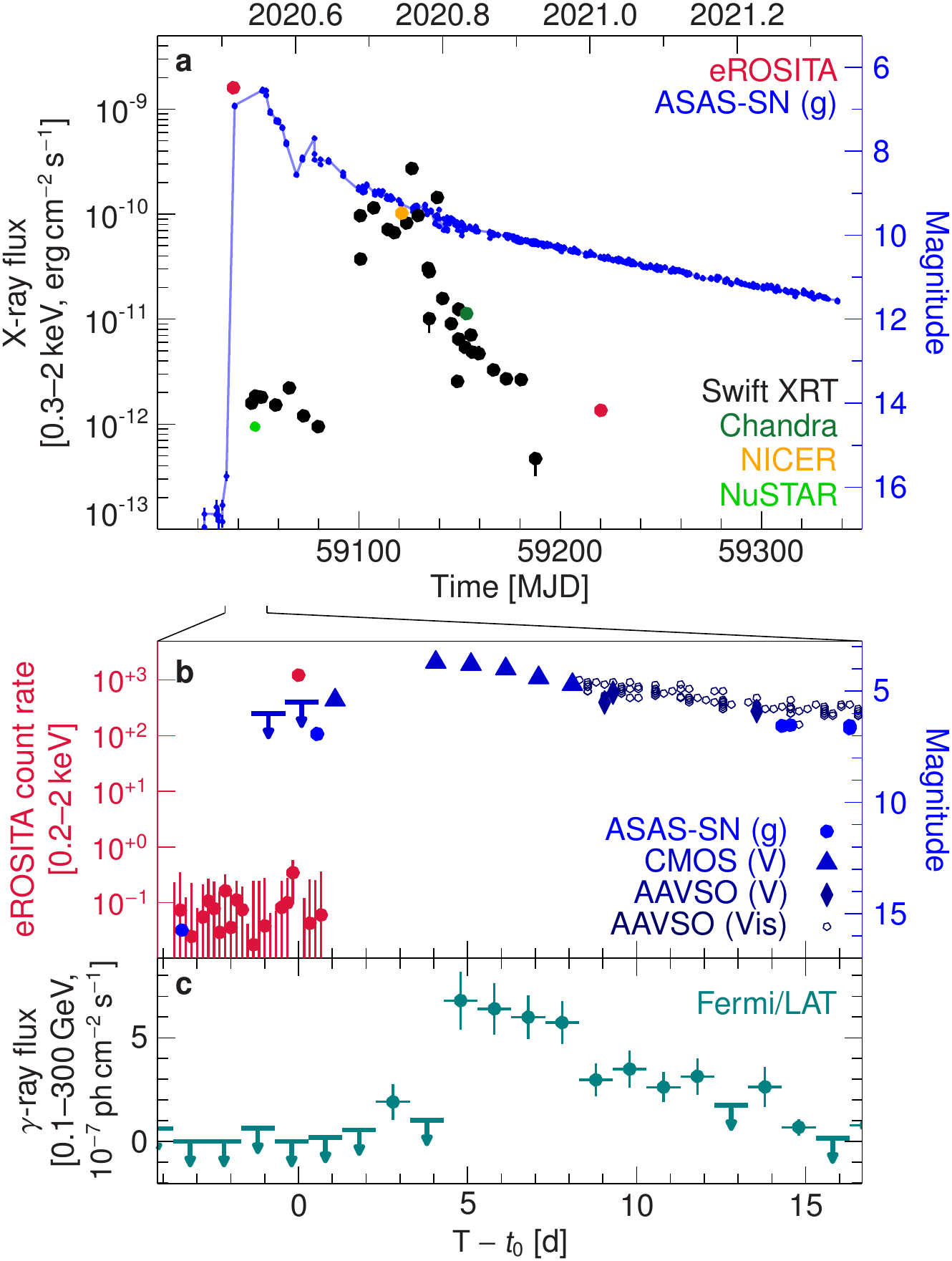}
  \caption{Multi-wavelength lightcurve of YZ~Ret. Error bars denote
    $1\sigma$ confidence levels. \textbf{a} Long-term evolution
    tracing the absorbed X-ray flux and the optical flux from the
    flash through the supersoft state using eROSITA, Swift, NICER, and
    Chandra data. The extrapolated NuSTAR flux\citep{Sokolovsky20a}
    is multiplied by 100 for visibility. \textbf{b} Short-term
    lightcurve before and after the X-ray flash showing the eROSITA
    count rate and the subsequent optical brightening. \textbf{c}
    Fermi/LAT lightcurve showing the $\gamma$-ray activity starting a
    few days after the flash.}
  \label{fig:master_lc}
\end{figure}

Figure~\ref{fig:master_lc} shows the combined multi-wavelength
lightcurve of YZ~Ret. At $t_0-3.5\,\mathrm{d}$, ASAS-SN monitoring
indicates the source close to the 16.4\,mag detection limit at
15.8\,mag. At $t_0+2.2$\,h an optical (V-band) non-detection with a
lower limit of 5.5\,mag was
reported\citep{McNaughtCBET4812,Sokolovsky20a}, followed by a fast
brightening at $t_0+11.3\,\mathrm{h}$ (2020-07-08 04:05:58 UT). The
nova reached a peak V-band brightness of
3.7\,mag\citep{McNaughtCBET4812} at $t_0+4.1\,\mathrm{d}$, making it
visible to the naked eye and the second brightest nova of the decade.
From this chronology of the events we conclude that eROSITA has
detected the X-ray ignition flash of a nova and that the X-ray flash
happened a few hours before the optical rise, as theoretically
predicted\citep{Krautter08a,Kato16a,Chomiuk21a}.

In other wavelengths, Fermi/LAT detected significant $\gamma$-ray
emission starting at $t_0+2.8$\,d (ref.\citep{Li20a}),
and a NuSTAR observation detected hard X-rays at $t_0+10.3$\,d (ref.\citep{Sokolovsky22a}). The hard emission is due to internal
shocks within the expanding shell. Starting approximately two months
after the X-ray flash, multiple missions\citep{Sokolovsky22a},
including eROSITA (Extended Data Fig.~\ref{fig:supersoftsource}), showed that
YZ~Ret had entered the supersoft state.

The progenitor of YZ~Ret is the known WD system MGAB-V207
\citep{Kilkenny15a}, which had a pre-eruption orbital period of
0.1324539(98)\,days\citep{schaefer21a}. Irregular variations in the
optical before the nova event suggest the system to be a VY~Scl type
cataclysmic variable\citep{Sokolovsky22a}. The nature and
composition of the WD is still unclear\citep{Sokolovsky22a}.

\section*{Spectral Analysis}

During the eROSITA detection of the flash, YZ~Ret was in the field of
view for 35.8\,s, with a count rate in excess of
1,000\,$\mathrm{counts}\,\mathrm{s}^{-1}$ (see
Fig.~\ref{fig:master_lc}b). Although the strong signal makes the
detection of the flash unambiguous, it complicates a more detailed
analysis of the properties of the nova: eROSITA's detectors are
severely affected by photon pile-up at such high count rates. The
nonlinear distortion of the spectral information requires careful
modeling of the response of the instrument to such a bright source. As
discussed in more detail in the Methods section, our approach includes
simulations using the same observing strategy as in the real
observation. The Simulation of X-ray Telescopes
(SIXTE)\citep{Dauser19a} software package is a generic Monte Carlo
toolkit for X-ray astronomical instrumentation and has been
particularly tailored to model pile-up in the eROSITA detectors. The
simulations allow us to perform a quantitative analysis and robustly
recover the basic properties of the source even when considering the
remaining systematic uncertainties. In the following, unless mentioned
otherwise, all uncertainties denote $3\sigma$ confidence levels.

\begin{figure}
  \centering
  \includegraphics[width=\linewidth]{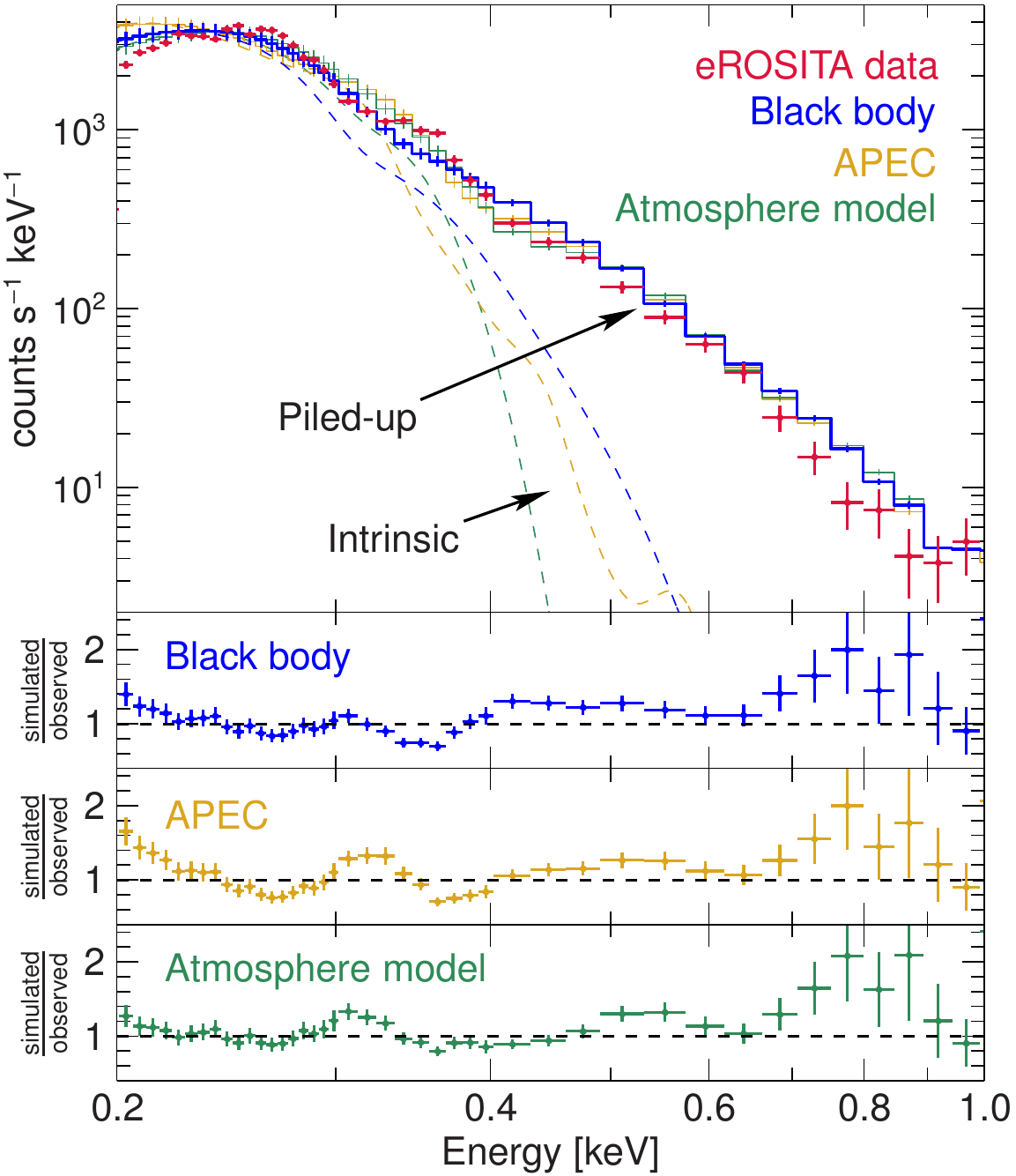}
  \caption{Comparison of measured and simulated spectra of the X-ray
    flash and the ratios between the best-fit models and the data.
    Solid lines show the piled-up model data, while dashed lines show
    the spectral shape for an observation without pile-up. Error bars
    are at the $1\sigma$ level and are shown assuming systematic
    uncertainties of 10\%. Model parameters are given in Extended Data
    Table~\ref{tab:fit_pars}.}
  \label{fig:spectrum}
\end{figure}

We investigate the data using three different models representative
for the range of expected spectral shapes, an empirical black body, an
optically thin collisional plasma (APEC)\citep{smith01a}, and a WD
atmosphere model\citep{Suleimanov13a,Suleimanov14a}. The best-fitting
models are shown in Fig.~\ref{fig:spectrum} and in the Extended Data
Table~\ref{tab:fit_pars}. For the black body model we find that the
spectrum can be best described at a 0.2--10\,keV absorbed flux of
$1.86^{+0.38}_{-0.23}\,\times
10^{-8}\,\mathrm{erg}\,\mathrm{cm}^{-2}\,\mathrm{s}^{-1}$ with a
temperature of $kT_\mathrm{BB}=28.2^{+0.9}_{-2.8}$\,eV
($3.27^{+0.11}_{-0.33}\times 10^5$\,K), where $k$ is the Boltzmann
constant. The foreground and internal absorption column to the source
is not very well known and the latter can also vary throughout the
outburst. Estimates of the Galactic equivalent hydrogen column are
$1\times 10^{19}\,\mathrm{cm}^{-2} \lesssim N_\mathrm{H} \lesssim
1.86\times 10^{20}\,\mathrm{cm}^{-2}$, and intrinsic absorption has
been estimated to be $\lesssim 1.5\times 10^{23}\,\mathrm{cm}^{-2}$ at
$t_0+10\,\mathrm{d}$ (refs.\citep{Izzo20a,Sokolovsky22a}).
Our black body fit constrains the equivalent hydrogen column density
to $N_\mathrm{H}<1.4\times 10^{20}\,\mathrm{cm}^{-2}$, indicating no
major intrinsic absorption during the X-ray flash. A comparison of the
measured and simulated lightcurve shows a possible decline during the
last few seconds of the observation (see Methods, Extended Data Fig.~\ref{fig:sixte_lightcurve}).

The best-fit atmosphere model has a 0.2--10\,keV absorbed flux of
$1.82^{+0.22}_{-0.28}\,\times
10^{-8}\,\mathrm{erg}\,\mathrm{cm}^{-2}\,\mathrm{s}^{-1}$ and a
temperature of $27.1^{+1.2}_{-0.5}\,\mathrm{eV}$, consistent with the
parameters of the black body model. In the fits we fixed
$N_\mathrm{H}=1\,\times 10^{19}\,\mathrm{cm}^{-2}$, i.e., we set
$N_\mathrm{H}$ to the lower limit of Galactic equivalent hydrogen
column estimates\citep{Izzo20a}. Within the systematic uncertainties
of the pile-up modeling, the black body and atmosphere models cannot
be further distinguished.

The residuals of the APEC fit are significantly worse, especially at
low energies, consistent with results that emission lines, presumably
due to shocks, typically emerge in the X-rays only days to months
after the eruption\citep{Chomiuk21a}. Similar to the later supersoft
phase, any shocked plasma emission is likely orders of magnitude
fainter than the supersoft emission. Therefore, we conclude that
eROSITA detected an optically thick thermal spectrum from the
photosphere.

\section*{The photosphere of YZ~Ret during the flash}
Assuming spherical emission and the Gaia distance of
$2.53\pm 0.26$\,kpc, the unabsorbed bolometric flux of the black body
model,
$2.6^{+1.4}_{-0.6}\times
10^{-7}\,\mathrm{erg}\,\mathrm{cm}^{-2}\,\mathrm{s}^{-1}$, corresponds
to a bolometric luminosity of
$(2.0\pm 1.2)\times 10^{38}\,\mathrm{erg}\,\mathrm{s}^{-1}$. This
luminosity is consistent with the theoretical prediction that the
source radiates at the Eddington luminosity during the ``fireball''
phase ($2 \times 10^{38}\,\mathrm{erg}\,\mathrm{s}^{-1}$ for a
canonical WD\citep{Kato15a,Kato16a}). At $t_0+4.1\,\mathrm{d}$ Nova
YZ~Ret was estimated\citep{Sokolovsky22a} to have a peak optical
bolometric luminosity of
$6.6\times 10^{38}\,\mathrm{erg}\,\mathrm{s}^{-1}$ (correcting the
distance from Gaia DR2 to Gaia EDR3). Since the high-energy luminosity
at that time is a factor of ${\sim}2,500$
fainter\citep{Sokolovsky22a}, we can assume that this luminosity
is representative of the total bolometric luminosity. The nova is
expected to evolve at approximately constant, Eddington-limited
bolometric luminosity, with the spectral peak moving from the X-rays
during the initial flash, to a peak in the optical at the maximum
expansion of the photosphere. Our derived luminosity is on the same
order of magnitude as at $t_0+4.1\,\mathrm{d}$, implying that the
total nova energetics are approximately conserved throughout the
outburst. The bolometric luminosity from the black body model during
the flash corresponds to a photospheric radius of
$50,000\pm 18,000\,\mathrm{km}$. The atmosphere model yields a
consistent radius of $37,000\pm 2,900\,\mathrm{km}$.

Theoretical work on nova outbursts shows that in the minutes after the
thermonuclear runaway the rise in bolometric luminosity occurs at
approximately constant radius\citep{JoseHernanz98a,hillman14}. Once
the energy has diffused from the bottom of the envelope to the
photosphere (about 5--10\,minutes after the onset of the runaway)
expansion starts immediately, and the envelope is
ejected\citep{aydi20} with velocities that can eventually reach up to
$6,000\,\mathrm{km}\,\mathrm{s}^{-1}$. Thus, the photosphere will
become orders of magnitude larger than a WD in a matter of minutes to
hours. Given that our derived photospheric radius is only a few times
larger than a typical WD radius
($3,500 \lesssim \mathrm{R}_\mathrm{WD}\lesssim
18,000$\,km)\citep{Bedard17a}, eROSITA detected the ``naked''
photosphere just after the released energy reached the surface of the
photosphere, before the main expansion of the envelope. At
$t_0+10.3$\,d the shocked region of YZ~Ret had expanded to an
estimated $<1.6$\,astronomical units\citep{Sokolovsky22a}.

The eROSITA detection provides further constraints on 
the temperature evolution of the photosphere during a complete nova
outburst. The effective temperature during the flash is expected to
peak in the range
$40\,\mathrm{eV}\lesssim kT \lesssim 100\,\mathrm{eV}$ and then
gradually decline to ${\sim}5$\,eV as the envelope expands over a
period of a few hours to days and the peak emission shifts to the
optical\citep{hillman14,Kato16a}. When the photosphere recedes back
closer to the WD surface during the supersoft state, the temperature
is expected to take a similar value as during the flash. The detailed
temperature profile depends on the core temperature, accretion rate,
and mass of the WD, but this overall pattern is ubiquitous for typical
novae\citep{hillman14}. The temperature during the eROSITA
observation, $kT\sim28$\,eV, is slightly below the expected peak
temperature, suggesting that the observation was in the gradual decay
phase of the flash. This interpretation is consistent with the
possible flux decrease in the last ${\sim}6$\,s of the observation.
During the supersoft state the temperature was ${\sim}30$\,eV
(Extended Data Fig.~\ref{fig:supersoftsource})\citep{Sokolovsky20b}, which
confirms that similar temperatures are measured during the flash and
the supersoft phase.

Theoretical studies also predict that the existence and duration of
the ignition flash correlates with the WD mass
\citep{hillman14,Kato17a}. Low-mass WDs with ${\sim}0.65\,M_\odot$ are
predicted to show only near-UV flashes, with durations of 5--10\,d,
while X-ray flashes are expected in moderate to high-mass WDs with
durations of 6--12\,h for a mass of $1.0\,M_\odot$ (ref.\citep{hillman14}).
The existence of the X-ray flash seen in YZ~Ret therefore implies a
relatively massive WD, which is confirmed by the short duration of the
flash, and consistent with the mass inferred from the atmosphere model
fits, $M_\mathrm{WD} = (0.98 \pm 0.23)\,M_\odot$, although this model
does not include the effects of the expansion of the atmosphere.

While UV emission has been detected for a few
novae\citep{cao12,pietsch07a}, searches for the X-ray
flash\citep{Morii16a,Kato16a} have so far been unsuccessful. The
luminosity limit from MAXI\citep{Morii16a} was significantly above
that expected from nova theory due to its 2\,keV low energy threshold,
while the Swift monitoring of the recurrent nova M31N
2008-12A\citep{Kato16a} was optimized for a nova with an hours-long
X-ray flash at a luminosity comparable to that of YZ Ret, and would
likely have missed shorter duration flashes of a similar brightness.

With the successful detection of the flash of YZ~Ret by SRG/eROSITA,
the existence of X-ray flashes has now been observationally confirmed.
Based on estimates for the Galactic nova rate\citep{kishalay2021},
eROSITA is expected to detect at most 1--2 of such events during its
4\,year survey phase. The eROSITA data provide the last link of the
predicted photospheric lightcurve evolution, since the initial UV
flashes and all post-maximum phases have been detected. Therefore, our
detection also adds the missing piece to measure the total nova
energetics and completes the whole picture of the photospheric
evolution of the thermonuclear runaway.

\section*{Methods}\label{sec:methods}
% 3000 words

\subsection*{eROSITA data reduction}\label{sec:pileup}
The eROSITA data analysis was performed with eSASS v201009, the
official analysis software of eROSITA. We exclude data of telescope
modules 5 and~7 that are contaminated by a light-leak. Images were
extracted using \texttt{evtool} and lightcurves and spectra using
\texttt{srctool}, selecting all valid patterns and a symmetric PSF
with no further event flagging. The data products of the flash are
extracted for the time range with a fractional exposure $>0.2$. We use
a source radius of $4.8'$ for the extraction of the X-ray flash data,
and $2'$ for the data taken during the supersoft phase half a year
later. In both cases we subtract a background spectrum, accumulated
from an annulus of radii $6'$ and $9'$ around the source position.
Throughout this paper, absorption modeling assumes the \texttt{wilm}
abundances for the interstellar medium\citep{Wilms00a} and
\texttt{vern} cross-sections\citep{Verner96a}.

\subsubsection*{Quantitative modeling of the spectra observed by eROSITA}
The major complexity in the modeling of the eROSITA data is the
so-called pile-up that is caused by the high photon flux of the
source\citep{Ballet99a,Davis01a,Dauser19a,Tamba22a}. When an
X-ray photon impacts a charged-coupled device (CCD) it creates a cloud
of electrons in the semiconductor. These electrons are then collected
in one or multiple detector pixels. During detector read-out, charge
below a certain threshold is discarded. The initial step of the event
reconstruction in the event processing pipelines reconstructs the
events from the measured charge distribution. Depending on the number
of pixels in which charge is detected during one read-out cycle,
events are classified into four patterns: singles, i.e., only one
pixel contains charge, doubles, triples, and quadruples. Because the
charge cloud size is energy-dependent, and because of thresholding
effects, the relative fraction of each of these pattern types depends
on energy.

Virtually all currently operating imaging X-ray telescopes are
designed to operate in single-photon mode, which allows direct
reconstruction of the energy of each incident photon. For a moderate
photon rate, the processing algorithm can correctly identify the
patterns and reconstruct the original photon energy from the summed
charges. For bright sources, however, multiple photons can hit the
same or neighboring pixels during one readout cycle. In the extreme
case the illuminated pixels produce extended charge distributions that
can be discarded in the event reconstruction, leading to a complete
loss of the signal, and a depression in brightness at the center of
the source (see Fig.~\ref{fig:flash_images}
and Extended Data Fig.~\ref{fig:sixte_lightcurve}). For many other cases, however, the
charge pattern deposited in the sensor by multiple photons cannot be
distinguished from that deposited by a single photon that has a higher
energy\citep{Ballet99a,Davis01a}. As a result the
reconstructed count rate is reduced and the spectral shape is hardened
(see Fig.~\ref{fig:spectrum}).

Because pile-up is strongly dependent on the source flux and spectral
shape, the typical forward-fitting approach of X-ray astronomy where
detector effects are modeled solely through a linear response matrix
\citep{lampton76} is not applicable. An additional problem in
the case of eROSITA's slew observations is that the pile-up is
time-dependent because vignetting degrades the point-spread function
(PSF) at the edge of the field-of-view (FOV). As a result, photons are
distributed over a larger number of pixels when the source enters and
leaves the FOV, resulting in less pile-up at the beginning and end of
the slew observation. This complexity also makes an excising of the
PSF core, which is a common mitigation approach for moderate pile-up
in pointed observations, unpractical for the eROSITA slew.

SIXTE\citep{Dauser19a}, the official eROSITA end-to-end simulator, is
capable of modeling the charge cloud, vignetting, and event
reconstruction during slew observations (see
refs.\citep{Townsley02a,Tamba22a} for similar
approaches). Comparison between existing on-ground and in-flight
eROSITA calibration measurements shows that the pre-flight PSF and
vignetting data reproduce the observations well. Since the pile-up
behavior of the detector is sensitively dependent on the charge cloud
size, the largest uncertainty in our simulation lies in the charge
cloud model.  SIXTE models the charge cloud as a 2D rotationally
symmetric Gaussian distribution, where the charge fraction in the
detector pixel $(n,m)$ is
$c_{n,m}=c_\mathrm{total}/(2\pi\sigma^2)\cdot \int_{x_n}^{x_{n+1}}
\int_{y_m}^{y_{m+1}} \exp
(-((x-x_\mathrm{i})^2+(y-y_\mathrm{i})^2)/(2\sigma^2))\,\mathrm{d}x\,\mathrm{d}y$.
$(x_\mathrm{i}, y_\mathrm{i})$ is the impact position of the photon,
$c_\mathrm{total}$ the total generated charge, and $\sigma$ the
standard deviation of the Gaussian charge cloud\citep{Dauser19a}.
Because the distribution of events in the real detector is
energy-dependent, $\sigma$ can be determined from the measured pattern
fractions. To this end, we derive the in-orbit energy-dependent
pattern fraction based on a large number of eROSITA slew observations
of bright Active Galactic Nuclei. Only events from the source regions
are included, as the particle background affects the low and high
energies (due to the abrupt reduction in effective area above the Au
M-edge of eROSITA's mirror system). Thus, we can empirically determine
$\sigma$ by minimizing the simulation against the calibration curve in
the 0.2--2\,keV range. The best-fit can be found at
$\sigma=9.8\,\mu\mathrm{m}$. The resulting pattern fractions are shown
in Extended Data Fig.~\ref{fig:pattern_fractions}, indicating that
SIXTE can reproduce the eROSITA in-flight pattern fractions at an
accuracy of a few percent in the energy range relevant here.

\subsubsection*{Spectral fitting with SIXTE}
For a given spectral model, which is characterized by a constant flux
and parameters such as the temperature, and modeling the foreground
absorption using interstellar medium abundances\citep{Wilms00a}, a
sufficiently large number of Monte Carlo realizations is averaged to
minimize the statistical noise in the simulated piled-up spectra. We
compute such model observations on a dense grid of spectral parameters
using SIMPUT v2.5.0, SIXTE v2.7.0, and the eROSITA instrument files
v1.9.1. For each grid point we simulate 1,000 slews (corresponding to
an effective exposure time of 31.5\,ks) using the attitude file of the
SRG spacecraft and the optical position of the source
($\alpha_\mathrm{J2000.0}=03\mathrm{h}\,58\mathrm{m}\,29.55\mathrm{s},
\delta_\mathrm{J2000.0}=-54^{\circ}\,46'\,41.23''$). We then compare
the simulated spectra against the measured data, varying the
parameters to minimize the $\chi^2$-statistics. This approach allows
us to derive the parameter uncertainties from the
$\Delta\chi^2$-contours (Extended Data Fig.~\ref{fig:chi2_landscape}). We estimate
that due to the uncertainties inherent to the charge cloud modeling,
systematic errors of $\sim$10\% will dominate the final error budget.
This systematic uncertainty is incorporated into the error
propagation.

The simulation is accurate enough to determine the black body or
atmosphere temperature to a relative uncertainty of ${\sim}10\%$ and
the flux to ${\sim}20\%$ (see Extended Data Table~\ref{tab:fit_pars}). We emphasize
that due to our fitting approach and remaining uncertainties in the
SIXTE pile-up model and other calibration data, the modeling is
dominated by systematic uncertainties. Our approach also allows us to
simulate the piled-up lightcurve and sky image, which are both very
similar to the measurement (Extended Data Fig.~\ref{fig:sixte_lightcurve}). The fact
that we can reproduce the piled-up image, lightcurve, and spectrum
with a reasonable set of parameters shows that SIXTE is indeed capable
of adequately modeling the very strong pile-up of this source.

\subsection*{Atmosphere model}
The plane-parallel atmosphere model assumes hydrostatic and local
thermodynamical equilibrium with abundances of hydrogen and helium
fixed to the solar value. We include the relevant heavier elements at
an abundance of half of the solar value. Photoionisation
cross-sections are computed for the ground\citep{Verner96a}, and
excited energy states\citep{Seaton94a}. In total, we consider
${\sim}25,000$ bound-bound transitions predominantly placed in the UV
and soft X-ray energy bands\citep{Dere97a}. The atmosphere model is
computed in the range $100\,\mathrm{kK}$--$1\,\mathrm{MK}$
($9\,\mathrm{eV}\lesssim kT_\mathrm{eff} \lesssim 90\,\mathrm{eV}$) at
a step size of 25\,kK (2.2\,eV). The second free parameter, the
surface gravity, is fixed to $\log g = \log g_\mathrm{Edd} + 0.1$
throughout this paper, where $\log g_\mathrm{Edd}$ is the surface
gravity corresponding to the Eddington luminosity at a given effective
temperature\citep{Suleimanov13a}. Since we do not consider the
line-driven wind arising in the upper layers of the atmospheres
\citep{Suleimanov13a}, the atmospheric parameters derived from the
spectral model should only be considered a first approximation. Note,
however, that the systematic uncertainties due to the treatment of
pile-up preclude a more detailed analysis using more physically
motivated models.

\subsection*{Multi-wavelength data}
\label{subsect:multiwavelength}

For the optical lightcurve of YZ~Ret, we use complementary V-band
($\lambda_\mathrm{eff}=5,448\,$\AA, $\Delta\lambda=890$\,\AA) and
g-band ($\lambda_\mathrm{eff}=4,639\,$\AA, $\Delta\lambda=1,280$\,\AA)
data from the All-Sky Automated Survey for Supernovae
(ASAS-SN)\citep{Kochanek17a}, V-band and visual (human eye,
${\sim}5,500\,$\AA) CCD data from the American Association of Variable
Star Observers (AAVSO).

The first X-ray measurement following the eROSITA observations was
taken only at $t_0+9\,\mathrm{d}$ with Swift XRT, when the nova
entered the supersoft state. We reduced the XRT spectra with HEASOFT
v6.26, using \texttt{xselect} v2.4k. A source extraction radius of
$1'$ for data taken in the photon counting (PC) mode and the same
background region as for eROSITA was used. We limited the PC and
windowed timing mode spectra to the 0.35--10\,keV energy range and
modeled the spectra with an absorbed black body of $\sim$30--40\,eV
and, when applicable, a thermal plasma model of
${\sim}4\,\mathrm{keV}$ (see also ref.\citep{Sokolovsky20b}). At
$t_0+9\,\mathrm{d}$ the X-ray flux of YZ~Ret was
$1.6^{+2.1}_{-1.3}\times
10^{-12}\,\mathrm{erg}\,\mathrm{cm}^{-2}\,\mathrm{s}^{-1}$
(0.3--10\,keV, see Fig.~\ref{fig:master_lc}, which also includes
complementary data from NICER\citep{Pei20a}, Chandra\citep{Drake20a}
and NuSTAR\citep{Sokolovsky20a}). The subsequent outburst behavior is
fully consistent with the picture of a nova in the supersoft source
state.

We extracted the 0.1--300\,GeV Fermi/LAT\citep{Atwood09a} daily
binned lightcurve using the Sciencetools (v1.2.23) and fermipy
(v0.20.0)\citep{WoodFermipy17}. The $15^\circ$ region of interest is
centered on the eROSITA position of YZ~Ret, with \texttt{SOURCE} class
events being selected and quality cuts being applied
(\texttt{DATA\_QUAL$>$0 \&\& LAT\_CONFIG==1}). Events with zenith
angles $\geq 90^\circ$ were discarded. We used the
\texttt{P8R3\_SOURCE\_V2} response for the extraction and
\texttt{gll\_iem\_v07} and \texttt{iso\_P8R3\_SOURCE\_V2\_v1} to model
the Galactic diffusion and the isotropic diffusion emission,
respectively. Datapoints with $\mathrm{TS}<9$ are given as upper
limits.

\subsection*{Expected rate of nova detections with eROSITA}
To estimate the expected number of nova detections during the eROSITA
survey, we use Monte Carlo simulations. Based on the Galactic
rate\citep{kishalay2021} of
$46.1\,\mathrm{novae}\,\mathrm{year}^{-1}$, we generate nova outburst
times and nova positions. These are uniformly distributed within
$18^\circ$ of the Galactic plane. We then use the as-flown attitude of
eROSITA for the year 2021 to determine the nova outbursts that would
have been detected during that year. In order to derive an upper limit
for the detection rate, we assume that all novae that pass through the
field of view are detected, i.e., we ignore the potentially severe
effects of Galactic absorption at lower Galactic latitudes, distance
effects, and the softening of the emitted radiation during the X-ray
flash. To accumulate statistics, we repeat this exercise for 10,000
times. We find that the probability of having one detection per year
is 8.6\% for X-ray flash durations of 3,600\,s, the probability of
having two detections per year is 0.4\%. The detection probability
drops to 2.6\% for one detection per year and a flash duration of
1,000\,s. Increasing the nova rate to 100 per year and assuming a
3,600\,s flash duration, the annual detection probability increases to
16\%. We therefore expect at most two detections of short X-ray
flashes for the four years of the eROSITA survey.

\backmatter

% Some journals require declarations to be submitted in a standardised format. Please check the Instructions for Authors of the journal to which you are submitting to see if you need to complete this section. If yes, your manuscript must contain the following sections under the heading `Declarations': Funding, Conflict of interest/Competing interests (check journal-specific guidelines for which heading to use), Ethics approval, Consent to participate, Consent for publication, Availability of data and materials, Code availability, Authors' contributions

% \bmhead{Supplementary information}
% If your article has accompanying supplementary file/s please state so here. 

\bmhead{Acknowledgments} This work is based on data from eROSITA, the
soft X-ray instrument aboard SRG, a joint Russian-German science
mission supported by the Russian Space Agency (Roskosmos), in the
interests of the Russian Academy of Sciences represented by its Space
Research Institute (IKI), and the Deutsches Zentrum f\"ur Luft- und
Raumfahrt (DLR). This work was supported by the Bundesministerium
f\"ur Forschung und Technologie under DLR grants 50\,QR\,1603,
50\,QR\,2103, and 50\,QR\,2104. GS acknowledges support from the
Spanish MINECO grant PID2020-117252GB-I00. VS thanks the Deutsche
Forschungsgemeinschaft (DFG) for financial support (WE1312/53-1). This
research has made use of ISIS functions (ISISscripts) provided by
ECAP/Remeis observatory and MIT
(http://www.sternwarte.uni-erlangen.de/isis/). We acknowledge with
thanks the variable star observations from the AAVSO International
Database contributed by observers worldwide and used in this research.
This version of the article has been accepted for publication, after
peer review but is not the Version of Record and does not reflect
post-acceptance improvements, or any corrections. The Version of
Record is available online at:
\url{http://dx.doi.org/10.1038/s41586-022-04635-y}. Use of this Accepted
Version is subject to the publisher’s Accepted Manuscript terms of use
\url{https://www.springernature.com/gp/open-research/policies/accepted-
 manuscript-terms}.

\bmhead{Author contributions}
R.A.\ identified the original event.
The eROSITA near real time analysis was developed by I.K., A.R., and P.W., the final data
extraction and reduction was performed by O.K., J.W., R.A., S.H., A.Ma., including
calibration information from K.D.. The pattern fraction analysis for SIXTE was
performed by  O.K., S.H., T.D., and K.D.. SIXTE is being developed by 
T.D., C.K., M.L., O.K., and J.W.. The interpretation of the result 
was done by O.K., J.W., G.S., A.Me., A.Ma., T.D., K.W., V.D., A.S., F.H., and R.A.. The 
white dwarf atmosphere models are provided by V.S., T.R., and K.W.. The manuscript
was written by O.K., J.W.. J.W. and A.S. acquired funding for this work.

\bmhead{Data Availability} 
SIMPUT files for the best-fit model and calibrated eROSITA products of the
observation are available from \url{https://erosita.mpe.mpg.de/specialreleases/}. 

\bmhead{Code Availability} The SIXTE code and eROSITA instrument files are
publicly available at
\url{https://www.sternwarte.uni-erlangen.de/research/sixte/}. The
eROSITA analysis software, eSASS, is available at
\url{https://erosita.mpe.mpg.de/}.

\bmhead{Competing Interests} The authors declare no competing
interests.

\bmhead{Material and Correspondence}
Requests for data and correspondence should be sent to O.~K\"onig or J.~Wilms. 

\clearpage
\setcounter{figure}{0}
\section*{Extended Data}

\begin{figure}[h]
  \centering
  \includegraphics[width=1\linewidth]{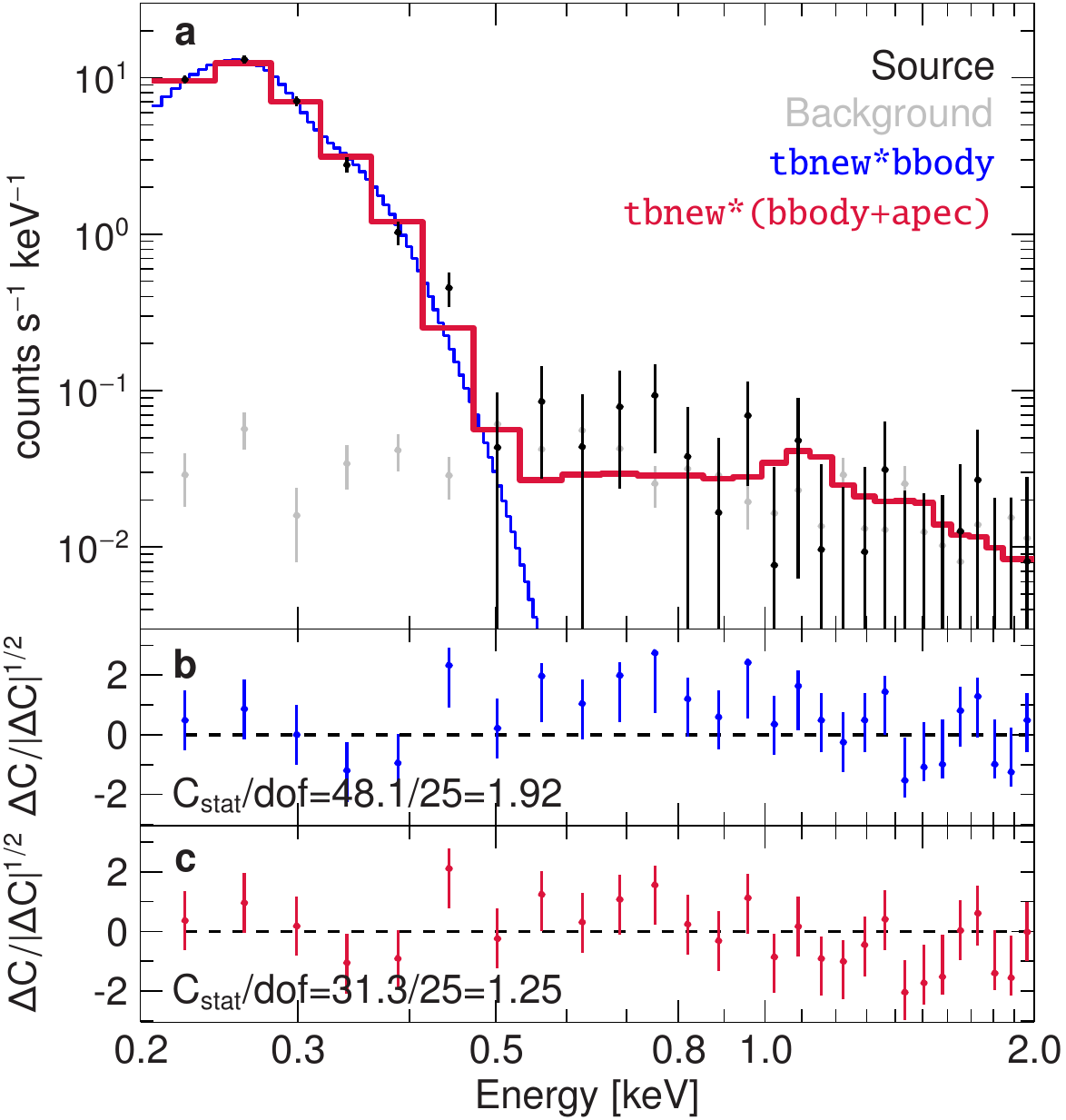}
  \caption{eROSITA spectrum of YZ~Reticuli taken during the supersoft
    state between $t_0+176.3$\,d and $t_0+186.3$\,d. The total
    exposure time is 640\,s, the spectrum is background subtracted.
    The spectrum can be described by a $20.7^{+0.7}_{-0.4}$\,eV black
    body under an equivalent hydrogen column density of
    $7.1^{+0.3}_{-0.9}\times 10^{20}\,\mathrm{cm}^{-2}$ and a
    0.3--2\,keV absorbed flux of
    $1.35(9) \times
    10^{-12}\,\mathrm{erg}\,\mathrm{cm}^{-2}\,\mathrm{s}^{-1}$. The
    reconstructed source position is $1.2''$ from the optical
    position, which makes source confusion very unlikely. The fit is
    based on Cash statistics\citep{Cash79a}, error bars are given at
    the $1\sigma$ confidence level. Panel \textbf{b} shows the
    residuals using only an absorbed black body model. While the
    statistics are formally better when including an additional APEC
    model, as shown in \textbf{c}, the data are consistent with the
    background level at energies $>$0.6\,keV.}
  \label{fig:supersoftsource}
\end{figure}

\begin{figure}
  \centering
  \includegraphics[width=\linewidth]{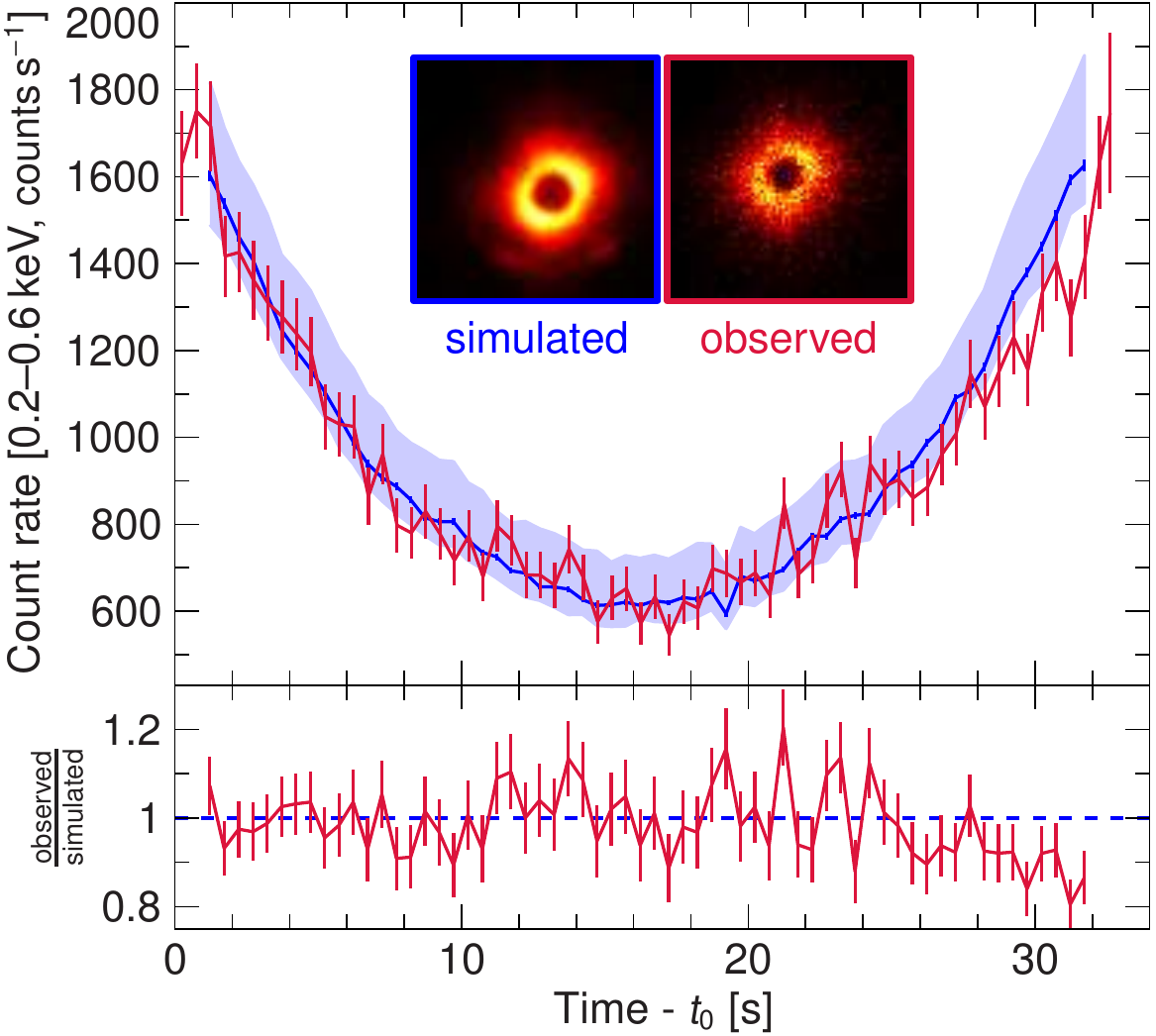}
  \caption{Comparison of the observed eROSITA slew lightcurve of the
    X-ray flash and the (averaged) simulation of a constant source
    with the best-fit black body parameters. The trough shape is due
    to pattern pile-up when the source passes the center of the FOV
    and vignetting. The last few seconds of the lightcurve show a
    possible decline in brightness. Error bars are at the $1\sigma$
    level, the blue shaded region indicates the $3\sigma$ flux
    uncertainty. The inset shows the observed source and (averaged)
    simulated image.}
  \label{fig:sixte_lightcurve}
\end{figure}

\begin{figure}
  \centering
  \includegraphics[width=1\linewidth]{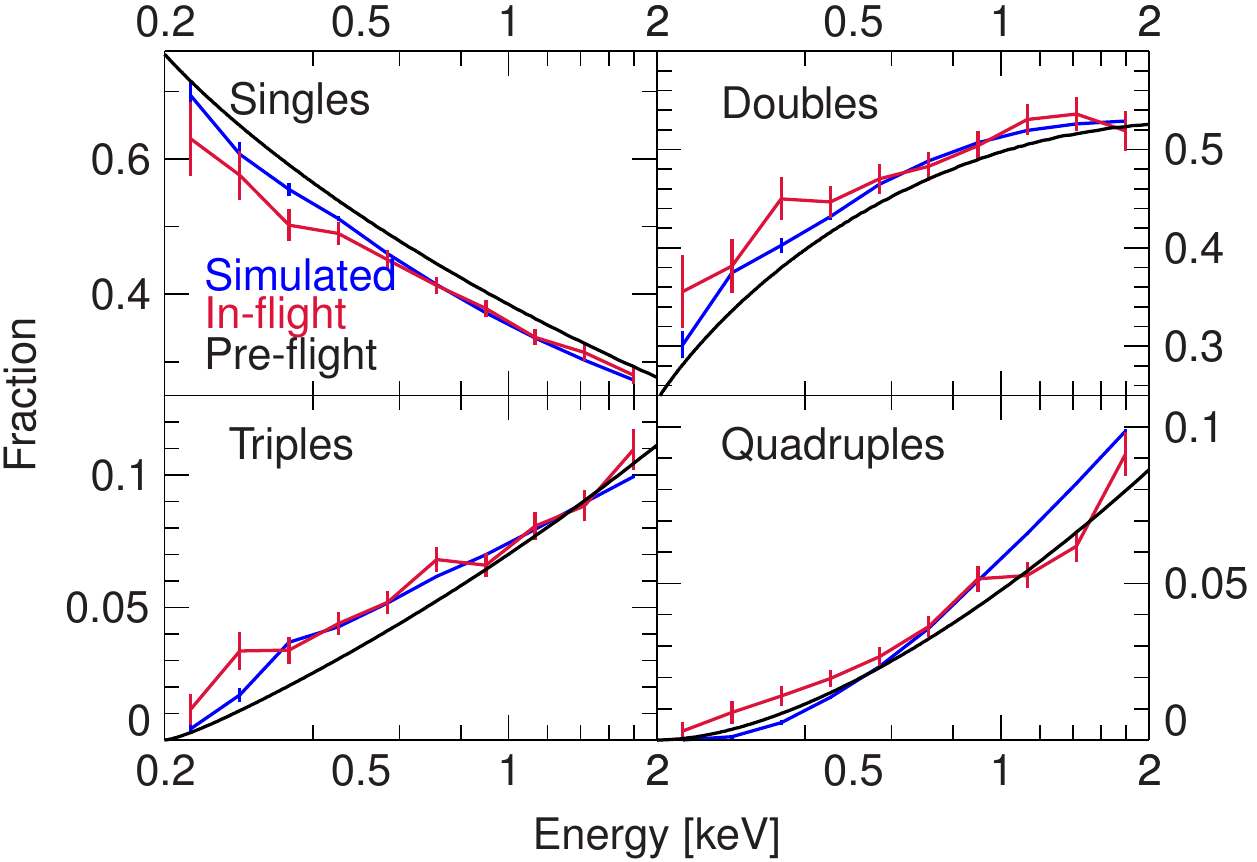}
  \caption{Comparison of simulated and measured pattern fractions in
    order to verify the pile-up model of SIXTE. The pre-flight pattern
    fractions from the TRoPIC prototype camera are shown for
    comparison\citep{Dennerl12a}. Error bars are at the $1\sigma$ level.}
  \label{fig:pattern_fractions}
\end{figure}

\begin{figure}
  \centering
  \includegraphics[width=\linewidth]{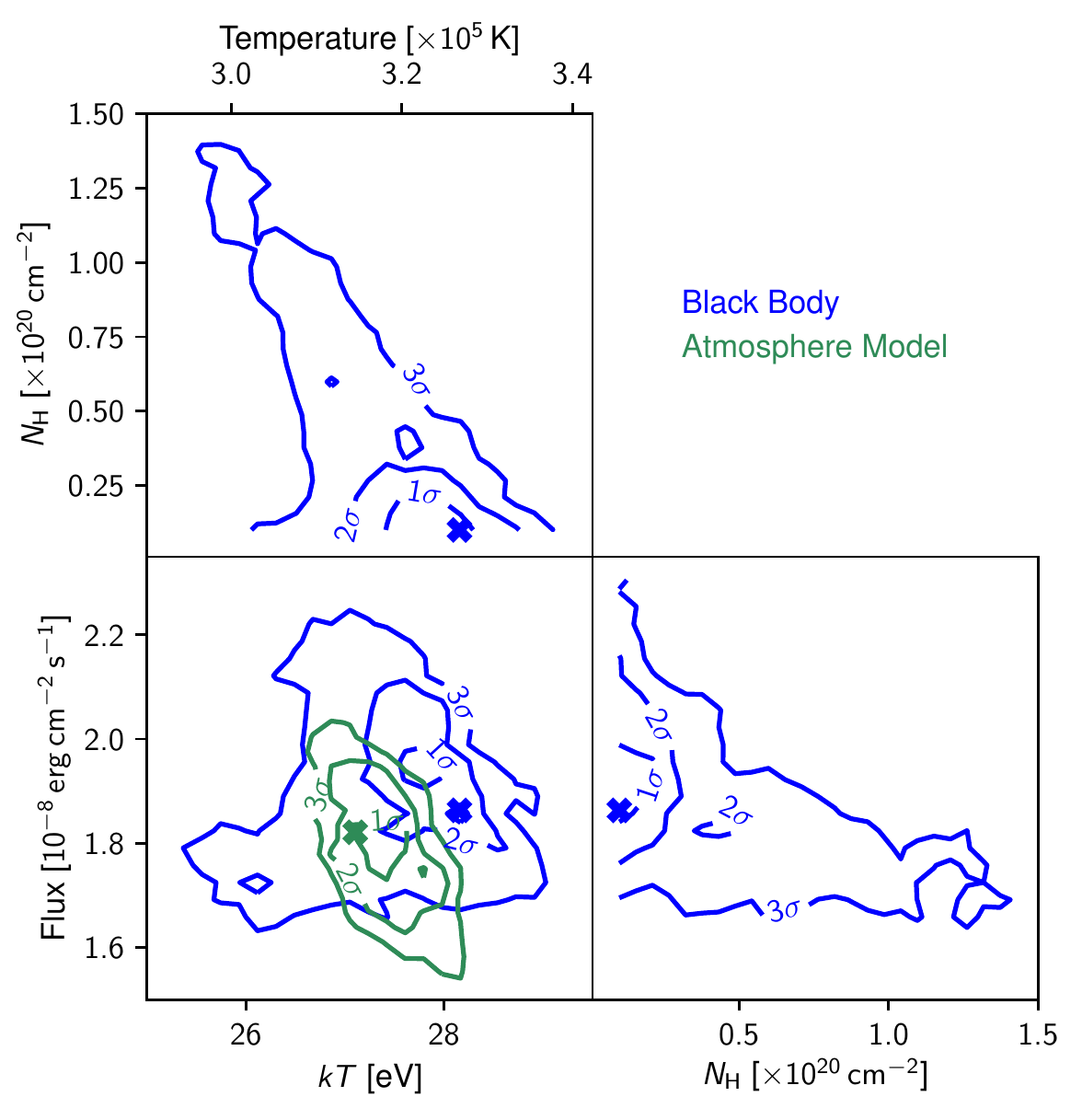}
  \caption{Parameter contours of the black body and atmosphere model
    fits. The contours show the $\Delta\chi^2$ between the averaged
    simulated spectra with respect to the observed spectrum. A
    systematic uncertainty of 10\% is assigned to each simulated
    spectrum. Contour lines give the $\Delta\chi^2$ values for 2
    degrees of freedom. The crosses denote the best-fit values, which
    are used in Fig.~\ref{fig:spectrum}.}
  \label{fig:chi2_landscape}
\end{figure}

\begin{table}%[h]
\begin{center}
\begin{minipage}{\textwidth}
  \caption{Best-fit models of the X-ray flash of YZ~Ret.}\label{tab:fit_pars}
\begin{tabular*}{\textwidth}{@{\extracolsep{\fill}}llll@{\extracolsep{\fill}}}
    \toprule%
    Model & \texttt{tbnew*bbody} & \texttt{tbnew*atmos} & \texttt{tbnew*apec} \\
    \midrule
    $kT$ [eV] & $28.2^{+0.9}_{-2.8}$ & $27.1^{+1.2}_{-0.5}$ & 25.9 \\
    Absorption\footnotemark[1] & $<1.4$ & 0.1 (fixed) & 1.7 \\
    Absorbed flux\footnotemark[2] & $1.86^{+0.38}_{-0.23}\times 10^{-8}$ & $1.82^{+0.22}_{-0.28}\times 10^{-8}$ & $2.5\times 10^{-8}$ \\
    $\chi^2_\mathrm{red.}$ & $228.3/(130-3)=1.80$ & $179.1/(130-2)=1.40$ & $255.0/(130-3)=2.01$ \\
    Luminosity [$\mathrm{erg}\,\mathrm{s}^{-1}$] & $2.0(1.2)\times 10^{38}$ & $0.98(22)\times 10^{38}$ & \\
    Radius [km] & $50,000\pm 18,000$ & $37,000\pm 2,900$ & \\ 
    Notes & & $\log \mathrm{g}=6.97\pm 0.17$ & Solar abundances \\
    \botrule
  \end{tabular*}
       Uncertainties are given at the $3\sigma$ confidence
    level for one parameter of interest.
  \footnotetext[1]{Equivalent hydrogen column density in units of $10^{20}$\,cm$^{-2}$.}
  \footnotetext[2]{Absorbed flux in units of
    $\mathrm{erg}\,\mathrm{cm}^{-2}\,\mathrm{s}^{-1}$ in
    0.2--10\,keV.}
\end{minipage}
\end{center}
\end{table}

\clearpage

\bibliographystyle{naturemag}
%% \bibliography{sn-bibliography}

\end{document}